\documentclass[12pt,a4paper]{article}   
\usepackage[dvips]{graphicx}
\usepackage{amsmath,amsfonts,amssymb}

\newcommand{\pn}{\par\noindent}
	  
\newcommand{\np}{\nonumber\\}
\newtheorem{thm}{Theorem}

\newtheorem{cj}{Conjecture}

\title{
Functional Relations in Stokes Multipliers and Solvable Models
related to $U_q(A^{(1)}_n)$
}
\author{ J. Suzuki\thanks{e-mail: sjsuzuk@ipc.shizuoka.ac.jp}\\
        \parbox{0.9\textwidth}{
        {\em
        \begin{center}
       Department of Physics, Faculty of Science\\
       Shizuoka University,\\
      Ohya 836, Shizuoka,\\
       Japan
        \end{center}
        }}
       }
\date{October 1999}
\begin{document}
\maketitle
\begin{abstract}
Recently, Dorey and Tateo have investigated 
 functional relations among Stokes multipliers for 
a Schr{\"o}dinger equation (second order differential equation)
with a polynomial potential term 
 in view of solvable models.
Here we extend their studies to a restricted case of 
$n+1-$th order linear differential equations. 
\end{abstract}
\clearpage
%+++++++++++++++++++++++++++++++++++++++++++++++++++++++ Sec1

\section{Introduction}\label{intro}
In remarkable papers\cite{DT1,DT2}, Dorey and Tateo find marvelous
connections between  a Schr{\"o}dinger equation
with a polynomial potential term and solvable models.
The success integrates two ingredients.
The exact  WKB analysis\cite{V1}-\cite{Takei} is on the one side.
The development in studies of solvable models 
 and of the solvability structure
in (perturbed) CFT \cite{Baxbook, BLZ1, BLZ2, FLS} 
via the thermodynamic Bethe ansatz\cite{AlZ, KP}
and nonlinear integral equations is on the other\cite{KBP}-\cite{SuzSpinS}.\pn
In view  of  the eigenvalue problem in quantum mechanics, 
${\cal H}\Psi = E \Psi$, 
 the quantity of interest is the spectral determinant,
${\rm det}({\cal H}-E)$.
For the potential term $x^{2M}$ with $M=$ integer,
 it coincides with the "vacuum expectation value" of
 a particular  member of a fusion transfer matrices in a certain field theory
possessing $U_q(\widehat{\mathfrak{sl}_2})$  \cite{DT1,JS}.
The  rest  of members 
in the  fusion hierarchy  are identified with 
the Stokes multipliers and their generalization \cite{DT2} in the
vacuum sector.
Thus the result 
provides a unified view of Stokes multipliers and spectral determinant.\pn
The  spectral determinant for  the wider class of potentials
$x^{2M}+\ell(\ell+1)/x^2$ with $M$ general can be treated within a framework using 
 Baxter's $Q-$ operator and the quantum Wronskian relation\cite{BLZnew}.

We extend a part of studies in \cite{DT2} to the higher order
differential equation case,
\begin{equation}
\partial^{n+1} y + (-1)^n (x^{m} + \lambda^{n+1}) y =0,
\label{difeq}
\end{equation}
with $m$ integer.\pn
The $\lambda=0$ case
 is essentially equivalent to the Turrittin equation
 of which solutions are calculated in terms of Meijier $G-$ functions.
There have been several results on this case \cite{Scheffe,  Turrittin,
Trji}.

We consider the Stokes multipliers associated with eq.(\ref{difeq}).
They satisfy a set of functional relations in the complex $\lambda$ plane.
We will show  that nontrivial solutions to relations are expressible by 
the quantum Jacobi-Trudi formula  which appears in
fusion transfer matrices  related to $U_q(A^{(1)}_n)$ 
\cite{BR,KNS1}.

We note the very recent result in \cite{DT3}  where 
the eigenvalue problem of eq.(\ref{difeq}) 
is argued for  $n=2$ case with $x \in[0,+\infty]$
though a related but a different  method .

%++++++++++++++++++++++++++++++++++++++++++++++++

\section{Asymptotic Expansion and  Stokes Coefficients}\label{survey}

We refer \cite{BOOK} to readers  for the background on the subject. 

Let us first discuss the asymptotic behavior of a slightly generalized
differential equation,
\begin{eqnarray}
& &\partial^{n+1} y + (-1)^n P(x) y =0  \np
& & P(x) = \sum_{j=0}^m a_j x^{m-j} 
\label{difp}
\end{eqnarray}
where $a_j$ are complex numbers and $a_0=1$.
$P(x)$ will be referred to as the potential term in analogy with
Schr{\"o}dinger equations.
\footnote{
In the case of singularity at infinity
of rank 1, i.e., $P(x)=p_0+p_1/x + \cdots$ in the above,
asymptotic solutions and numerical algorithms are discussed in
\cite{Olver} for a wider class of higher order differential equations.
}
The factor $(-1)^n$ is not essential.
It can be adsorbed into re-definition of the angle of $x$ and
$a_{j \ge 1}$.
For later convenience, we include this factor throughout this report.

Now that $x=\infty$ is an irregular singular point of the equation,
 analytic properties of solutions depend on sectors in the complex $x$ plane.
Let ${\cal S}_k$ be a sector in the plane satisfying
$$
|{\rm arg} x -k \theta  | \le \frac{\theta}{2}  
$$
for $ x \in {\cal S}_k$, where $\theta= \frac{2 \pi}{m+n+1}$.
We then study the asymptotic behavior of a subdominant solution in 
${\cal S}_0$.
Following \cite{HS,Sbook}, we define $b_h (h=1, 2, \cdots)$ by the relation,
$$
(1+ \sum_{k=1}^{m} a_k x^{-k}) ^{1/(n+1)} = 1+\sum_{h=1}^{\infty} b_h x^{-h}.
$$
A key function $E(x, {\bf a})$ is determined  by $b_h$,
\begin{eqnarray*}
E(x,{\bf a}) &:=&\int
     (1+\sum_{h=1}^{h_m} b_h x^{-h}) x^{m/(n+1)} dx  \\
        &=& \frac{n+1}{m+n+1} x^{(m+n+1)/(n+1)}+
		\sum_{h=1}^{h_m} \frac{b_h}{\frac{m}{n+1}-h+1} x^{m/(n+1)+1-h}
\end{eqnarray*}
where $h_m = N$ for $m=N(n+1) -j , (j=1, \cdots, n)$. 
Here ${\bf a}$ stands for $(a_1,a_2, \cdots, a_m)$

In addition, we introduce an exponent $\nu_m$ by
\begin{equation}
\nu_m=
\begin{array}{rl}
 \frac{nm}{2},&
           \quad \mbox{for $m \ne 0$ mod $n+1$}  \\
  \frac{nm}{2}+(n+1) b_{h_m+1},&
      \quad \mbox{for $m = 0$ mod $n+1$ }.\\
\end{array} 
\end{equation}
Then we have a theorem,
\begin{thm}
In  ${\cal S}_0$,  there exists a subdominant 
solution $y(x, {\bf a})$  to eq.(\ref{difp}) 
which admits the following asymptotics,
\begin{eqnarray}
y(x, {\bf a}) &\sim& x^{-\nu_m/(n+1)} e^{-E(x,{\bf a})}, \np
\partial^j y(x, {\bf a}) &\sim& x^{(jm-\nu_m)/(n+1)} e^{-E(x,{\bf a})},
\quad j \ge 1.
\label{asy}
\end{eqnarray}
\end{thm}

The range  of the existence of the asymptotic expansion (\ref{asy}) 
(forgetting the subdominance) actually extends over  ${\cal S}_0$.
The precise determination of the range, however, 
 poses a  nontrivial problem
even for $n=1$ case.  See elaborate discussions in sections 9 and 10 of \cite{Sbook}. 
The straightforward extension of the argument may lead to the conclusion that  
$y(x, {\bf a})$ admits the asymptotic expansion  (\ref{asy})  in an open sector,
\begin{equation}
{\rm arg }x < \frac{n+2}{m+n+1} \pi.
\label{open}
\end{equation}
See the appendix.

The intriguing feature in the differential equation (\ref{difp}) is
the following symmetry in rotating the $x-$ plane.
\begin{thm} 
Denote the above solution by $y(x, {\bf a})$. Then
\begin{equation}
y_k(x, {\bf a}) :=
y(x q^{-k}, G^{(k)}({\bf a})) q^{nk/2}
\label{symm}
\end{equation}
is also a solution to eq.(\ref{difp}).
\end{thm}
The  parameter $q$ signifies $\exp(i\theta) =\exp(i\frac{2\pi}{m+n+1})$.
The operation $G^{(k)}({\bf a})$ is defined by 
$G^{(k)}({\bf a})=G^{(1)}(G^{(k-1)}({\bf a}) ),
k \ge 2$ 
and 
$G^{(1)}({\bf a})=(a_1/q, a_2/q^2, \cdots  a_m/q^m)$.

A fundamental system of solutions (FSS) in ${\cal S}_k$ is formed by
$(y_k, y_{k+1}, \cdots, y_{k+n})$.
This is shown for the Turrittin equation in \cite{Trji}.
For general, it is not simple to show this directly.
We use the range of the validity of asymptotic expansion in  eq.(\ref{open}) and
 justify the linear independence  as follows. \pn

We introduce a $(n+1) \times (n+1)$ matrix $\Phi_k(x)$
and the Wronskian  $W_k:={\hbox {\rm det}} \Phi_k(x)$
\begin{equation}
\Phi_k(x):=
\begin{pmatrix}
y_k,            & y_{k+1},           & \cdots,& y_{k+n}\\
\partial y_k,   & \partial y_{k+1},  & \cdots,&\partial y_{k+n}\\
\vdots          &                    &        &\vdots \\
\partial^n y_k, & \partial^n y_{k+1},& \cdots,&\partial^n y_{k+n}
\end{pmatrix}.
\end{equation}
Then eq.(\ref{open})  implies that  $y_{k+j}, \, j=0, \cdots, n$
has the asymptotic expansion (\ref{symm}) 
in the closed sector ${\cal S}_{k+(n+1)/2} \cup {\cal S}_{k+(n-1)/2} $.
As $W_k$ is constant in $x$, 
one then verifies the linear independence of these solutions by
using the asymptotic expansions (\ref{asy}) and (\ref{symm})
in ${\cal S}_{k+(n+1)/2} \cup {\cal S}_{k+(n-1)/2} $. 
See discussions in \cite{Sbook} for $n=1$ and \cite{DT3} for $n=2$

We stress that the argument below depends on the existence of
$\Phi_k(x)$ as FSS, rather than (\ref{open}).

A matrix of Stokes multipliers $S^{(1)}_k$ connects FFS of ${\cal S}_{k}$
and ${\cal S}_{k+1}$

\begin{equation}
\Phi_{k}(x) = \Phi_{k+1}(x)S^{(1)}_{k}.
\label{stokes0}
\end{equation}

The linear independence of solutions fixes $S_k$ in the following form,

\begin{equation}
S^{(1)}_k =
\begin{pmatrix}
\tau^{(1)}_k,  & 1, & 0,&  0,&  \cdots, & 0\\
\tau^{(2)}_k,  & 0, & 1,&  0,&  \cdots, & 0\\
      \vdots   &    &    &  &          & \vdots \\
\tau^{(n)}_k,  & 0, & 0,&  0,&  \cdots, & 1 \\
\tau^{(n+1)}_k,& 0, & 0,&  0,& \cdots,  & 0
\end{pmatrix}.
\label{sk0form}
\end{equation}

By Cramer's formula, one represents $\tau^{(j)}_k$ as

\begin{equation}
\tau^{(j)}_k=\frac{1}{W_k} {\rm  det }
\begin{pmatrix}
y_{k+1},& y_{k+2}, &\cdots, & y_k, &  \cdots, & y_{k+n+1}\\
\vdots  &          &        &      &          & \vdots \\
\partial^{n} y_{k+1},&\partial^{n} y_{k+2}, &\cdots,& \partial^{n}y_k,& 
 \cdots, & \partial^{n}y_{k+n+1}
\end{pmatrix}.
\label{taujform}
\end{equation}
The column vector, $(y_k, \partial y_k, \cdots, \partial^{n}y_k)$ is inserted
in the $j-$th column in the matrix of the denominator.
Especially, 
\begin{equation}
\tau^{(n+1)}_k = (-1)^n W_{k+1}/W_k.
\label{taun}
\end{equation}

Then we restrict ourselves to the case of our interest (\ref{difeq}).
We find that  $b_{h_m+1}=0$ and thus 
$\nu_m=nm/2$ for any $m$.
Under the operation of $G^{(1)}$,  $G^{(1)}(a_m) =a_m q^{-m} =a_m q^{n+1}$.
As $a_m=\lambda^{n+1}$, this means   $G^{(1)}(\lambda) =\lambda q$.
Thus a function in the $k-$ th sector has an argument  $\lambda q^k$. \pn
%
%Notice that the apparent symmetry of eq.(\ref{difeq}) under
%$\lambda \rightarrow \lambda \exp(i\frac{2\pi}{n+1})$ is not
%inherited to solutions and Stokes multipliers.
%
%The extreme limit $x \rightarrow 0$ may convince us that
%the above manipulation sends one WKB solution to another. 

%
%------------------------------------------------
%
\section{Recursion relations }
By the commensurability of the cone angle of ${\cal S}_{k}$ to 2$\pi$,
the product of successive $m+n+1$ Stokes matrices 
must be a unit matrix  \cite{HS} modulo sign factor due to the normalization
in (\ref{symm}).
\begin{equation}
S^{(1)}_{m+n} S^{(1)}_{m+n-1} \cdots S^{(1)}_0 = (-1)^n I.
\label{prod1}
\end{equation}
Regarding as a function of $\lambda$, $S^{(1)}_k = S^{(1)}(\lambda q^k)$.
Thus eq.(\ref{prod1}) reads,
\begin{equation}
S^{(1)}(\lambda q^{m+n}) S^{(1)}(\lambda q^{m+n-1}) \cdots S^{(1)}(\lambda) = (-1)^n I.
\label{prod2}
\end{equation}
The same relation can be  recapitulated in terms of 
 a generalized Stokes matrix 
 $S_k^{(\ell)}$
connecting $\Phi_k$ and $\Phi_{k+\ell}$ \cite{DT2,BOOK},
\begin{equation}
\Phi_k = \Phi_{k+\ell} S_k^{(\ell)}.
\label{generalstokes}
\end{equation}
Similar to the above, $S_k^{(\ell)}= S^{(\ell)}(\lambda q^k)$ as a
function of $\lambda$.\pn
Eq. (\ref{prod2}) is obviously rewritten as, 
\begin{equation}
S^{(m+n+1)}(\lambda) = (-1)^n I.
\label{mostfusion}
\end{equation}
By the above definition, we have two equivalent recursion relations
\begin{eqnarray}
S^{(\ell+1)}(\lambda) &=& S^{(1)}(\lambda q^{\ell}) S^{(\ell)}(\lambda) \label{rec1}\\
                      &=& S^{(\ell)}(\lambda q)S^{(1)}(\lambda) \label{rec2}.
\end{eqnarray}
We denote the $(i,j)$ component of $S^{(\ell)}(\lambda)$ by
$S^{(\ell)}_{i,j}(\lambda)$. 

The main problems in the present report are 
\begin{description}
\item{(1)}  the expression of 
$S^{(\ell)}_{i,j}(\lambda)$ in terms of $\tau^{(a)}(\lambda)$s 
\item{(2)} restrictions imposed on  $\tau^{(a)}(\lambda)$s by  
functional relations among them.
\end{description}

The first recursion relation (\ref{rec1}) imposes the relations,
\begin{equation}
S^{(\ell+1)}_{k,j}(\lambda) =
\tau^{(k)}(q^{\ell}\lambda)S^{(\ell)}_{1,j}(\lambda)+
S^{(\ell)}_{k+1, j}(\lambda) \quad(1\le k\le n+1) \label{rec1c1} 
\end{equation}
where we formally put $S^{(\ell)}_{n+2, j}(\lambda) =0 $

The second recursion relation (\ref{rec2}) yields,
\begin{eqnarray}
S^{(\ell+1)}_{k,1}(\lambda) &=&
 \tau^{(1)}(\lambda)S^{(\ell)}_{k,1}(\lambda q)+
 \tau^{(2)}(\lambda)S^{(\ell)}_{k,2}(\lambda q)+ \np
 & &\cdots + \tau^{(n+1)}(\lambda)S^{(\ell)}_{k,n+1}(\lambda q) \label{rec2c1}
 \end{eqnarray}
and
\begin{equation}
S^{(\ell+1)}_{i,j+1}(\lambda)=S^{(\ell)}_{i,j}(q\lambda) \quad 
(1\le i \le n+1, 1\le j \le n).
\label{red2c2}
\end{equation}
The latter equation leads to
\begin{equation}
S^{(\ell)}_{i,j+1}(\lambda)=S^{(\ell-j)}_{i,1}(q^j \lambda) \quad \ell \ge j.
\label{red2c2d}
\end{equation}
For  $\ell < j$, 
\begin{equation}
S^{(\ell+1)}_{i,j}(\lambda)=S^{(1)}_{i,j-\ell}(q^{\ell} \lambda),
\label{red2c2e}
\end{equation}
and the rhs is given directly in terms of $\tau^{(a)}$.
Thus we regard only $S^{(\ell)}_{k,1}$ as nontrivial elements.

The eq. (\ref{rec2c1}) can be rewritten such that it contains only
$S^{(\ell)}_{k,1}$ in the lhs.
Let us write explicitly for $k=1$,
\begin{eqnarray}
S^{(\ell+1)}_{1,1}(\lambda) &=&
 \tau^{(1)}(\lambda)S^{(\ell)}_{1,1}(\lambda q)+
 \tau^{(2)}(\lambda)S^{(\ell-1)}_{1,1}(\lambda q^2)+ \np
 & &\cdots + \tau^{(n+1)}(\lambda)S^{(\ell-n)}_{1,1}(\lambda q^{n+1}).
\label{rec2c1d}
 \end{eqnarray}

We remark properties of $S^{(\ell)}_{1,1}(\lambda)$ from
linear independency of FSS.
The (1,1) component of eq.(\ref{generalstokes}) for $\ell=m+j, 1\le j \le n$
reads,
$$
y_k = S^{(m+j)}_{1,1} y_{k+m+j} + \cdots.
$$
 $\Phi_{k+m+j}$ consists of $(y_{k+m+j}, y_{k+m+j+1}, 
\cdots, y_{k+m+n+j})$. 
As $y_{k+m+n+j}=(-1)^n y_{k+j-1}$, it contains
$(-1)^n y_k$.
Thus the linear independency of solutions concludes at least
\begin{equation}
S^{(m+j)}_{1,1} =0, \quad 1\le j\le n.
\label{szero}
\end{equation}
We also note a trivial relation, $S^{(m+n+1)}_{1,1} = (-1)^n $.

In the next section, we will show that  a set of solutions to 
the above recursion relations can be neatly 
represented by the quantum Jacobi-Trudi formula.

%-----------------------------------------------------
%
\section{Quantum Jacobi-Trudi formula in Stokes multipliers}

The recursion relation (\ref{rec2c1d}) can be successively solved for
given initial conditions $\tau^{(1)}(\lambda), \cdots,\tau^{(n)}(\lambda)$
(forward-propagation).
Conversely, $\tau^{(a)}(\lambda)$ is expressible by these solutions
(back-propagation).

To represent these, we employ notations useful in later discussions,
\begin{eqnarray}
T^{(1)}_{\ell}(\lambda) &=& S^{(\ell)}_{1,1}(\lambda q^{-(\ell-1)/2}) \label{eq1} \\  
T^{(a)}_{1} (\lambda)&=& (-1)^{a+1} \tau^{(a)} (\lambda  q^{-(a-1)/2}) \quad
( =(-1)^{a+1} S^{(1)}_{a,1}(\lambda  q^{-(a-1)/2})   ).  \label{eq2}
\end{eqnarray}

%Conveniently, we extend the index  $\ell$ to non-positive values with
%$ T^{(1)}_{\ell}(\lambda) =0, \ell<0$ and $ T^{(1)}_{0}(\lambda) =0, \ell<0$.

We first state the result for the forward-propagation problem,
\begin{equation}
T^{(1)}_{\ell+1}(\lambda)={\rm det} 
\Bigl( T^{(1-i+j)}_1(\lambda q^{(\ell+2-i-j)/2}) \Bigr ) _{1\le,i,j \le \ell+1}.
\label{qJT1}
\end{equation}
The rhs includes the formal extension of $T^{(a)}_1(\lambda)$, 
and we understand $T^{(a)}_1(\lambda)=0$ 
for $a \ge n+2$ or $a <0$ and  $T^{(0)}_1(\lambda)=1$ .

The proof is easy by the induction on $\ell$.
The case $\ell=0$ is trivial by definition.
Assume  that eq.(\ref{eq1}) is valid up to $\ell$.
Expanding the determinant with respect to the last column,
we have
\begin{eqnarray}
T^{(1)}_{\ell+1}(\lambda)&=&T^{(1)}_1(\lambda q^{-\ell/2}) T^{(1)}_{\ell}(\lambda q^{1/2})
- T^{(2)}_1(\lambda q^{-(\ell-1)/2}) T^{(1)}_{\ell-1}(\lambda q)  \np
& &+ \cdots + (-1)^{a+1} T^{(a)}_1(\lambda q^{-(\ell+1-a)/2}) 
     T^{(1)}_{\ell+1-a}(\lambda q^{a/2}) +\cdots
\label{expand1}
\end{eqnarray}
Here we set 
 $ T^{(1)}_{\ell}(\lambda) =0, \ell<0$ and $ T^{(1)}_{0}(\lambda) =1$.

By eqs. (\ref{eq1}), (\ref{eq2}) and from the assumption of the induction , 
the lhs is 
equal to that in eq.(\ref{rec2c1d}) after 
$\lambda \rightarrow \lambda q^{-\ell/2}$.
Then the validity of eq.(\ref{eq1}) carries over into $\ell+1$, which completes 
the proof.

Next consider the back-propagation problem.
We regard the coupled equations  (\ref{rec2c1d}) for $\ell=0,1,2,\cdots, n$,
 as a set of linear equations for $\tau^{(a)}(\lambda)$
(or $T^{(a)}_1(\lambda)$).
The Cramer's theorem is again applicable.
Taking account of the fact that  the determinant of the coefficient matrix  is  unity,
we have,
\begin{equation}
T^{(a)}_{1}(\lambda)={\rm det} 
\Bigl( T^{(1)}_{1-i+j}(\lambda q^{(i+j-a-1)/2} )     \Bigr ) _{1\le,i,j \le a}.
\label{qJT2}
\end{equation}
It is natural to define $T^{(0)}_{1}(\lambda)=1$ from this expression.

We notice the similarity of eqs. (\ref{qJT1})  and (\ref{qJT2}) 
to the Jacobi-Trudi formula for Schur functions \cite{MacDo}.
It states that any complex Schur function associated to a skew Young diagram
can be represented in terms of a determinant of a matrix whose elements are
given by elementary Schur functions.

We find a quite parallel result in the present problem.
Thanks to the argument around eqs.(\ref{red2c2d}) and (\ref{red2c2e}),
only $S^{(\ell)}_{k,1}(\lambda)$  is of our interest.
This turns out to have a compact expression in a similar manner to 
the Jacobi-Trudi formula, which we call the quantum  Jacobi-Trudi formula.

To state this, we prepare some notations.
By $\mu$, we mean a Young diagram $(\mu_1, \mu_2, \cdots)$
and by $\mu'$, its transpose.
Consider a  skew Young table of the shape $\mu/\eta=(\mu_1-\eta_1,
\mu_2-\eta_2, \cdots)$ such that $\mu_i \ge \eta_i$.
We define a quantity associated to  $\mu/\eta$,
\begin{equation}
T_{\mu/\eta}(\lambda) 
  :={\rm det }_{1\le j,k\le \mu_1'} 
         ( T^{(1)}_{\mu_j-\eta_k-j+k} 
            (\lambda q^{-(\mu_1' -\mu_1+\mu_j+\eta_k-j-k+1)/2}) ).
\label{qJTsym}
\end{equation}
It contains eq.(\ref{qJT2}) as a special case
$\mu=(1,1,\cdots,1)$ and $\eta=\phi$.

A generalization of eq.(\ref{qJT1}) also exists.
To show this, consider two matrices $H$ and $E$  with entries 
\begin{eqnarray*}
H_{i,j}&:=&(-1)^{i-j} T^{(1)}_{i-j}(\lambda q^{-(i+j)/2}) ,  \\
E_{i,j}&:= &T^{(i-j)}_{1}(\lambda q^{-(i+j)/2}).
\end{eqnarray*}
Clearly, ${\rm det} H={\rm det} E=1$ as 
they are lower triangular  with all diagonal elements unity.
The expansion of eq.(\ref{qJT1}) with respect to the 1st column leads to
\begin{eqnarray*}
0&=&-T^{(1)}_{\ell+1}(\lambda)+T^{(1)}_1(\lambda q^{\ell/2}) 
T^{(1)}_{\ell}(\lambda q^{-1/2})
- T^{(2)}_1(\lambda q^{(\ell-1)/2}) T^{(1)}_{\ell-1}(\lambda q^{-1})  \np
& &+ \cdots + (-1)^{a+1} T^{(a)}_1(\lambda q^{(\ell+1-a)/2}) 
     T^{(1)}_{\ell+1-a}(\lambda q^{-a/2}) +\cdots
\end{eqnarray*}
which is equivalent to
$$
\delta_{i,j}= \sum_{k}  (-1)^{i-k} T^{(1)}_{i-k}(\lambda q^{-(i+k)/2} )
T^{(k-j)}_1 (\lambda q^{-(k+j)/2}).
$$
Hence $E$ and $H$ are inverse each other.

These two facts are sufficient for 
the second representation of $T_{\mu/\eta}$  \cite{BR,MacDo, KOS},
\begin{equation}
T_{\mu/\eta}(\lambda) 
=  {\rm det }_{1\le j,k\le \mu_1} 
         ( T^{(\mu'_j-\eta'_k-j+k)}_{1} 
            (\lambda q^{-(\mu_1' -\mu_1-\mu'_j-\eta'_k+j+k-1)/2}) ).     
\label{qJTasym}
\end{equation}
 Eqs. (\ref{qJTsym}) and (\ref{qJTasym}) 
coincide  with the Jacobi-Trudi formula by dropping the $\lambda$ dependencies.

We have the following statement on nontrivial Stokes multipliers.
\begin{thm}
The solution to the recursion relation (\ref{rec1c1}) for $j=1$ is given by
\begin{equation}
S^{(\ell)}_{k,1} (\lambda) = (-1)^{k+1} T_{\mu}(q^{(\ell+k-2)/2} \lambda),
\label{skew}
\end{equation}
where  $\mu$ is  a Young diagram of the hook shape
$\mu=(\ell,1,\cdots, 1)$ with
 height  $k$.
\end{thm}

The proof is as follows.
In terms of $T$, we need to show
$$
T_{\mu'}(\lambda) +T_{\mu"}(\lambda)=
 T^{(k)}_1(\lambda q^{\ell/2}) T^{(1)}_{\ell}(\lambda q^{-k/2}), 
$$
where $\mu' (\mu" )$ is  also the hook Young diagram 
$\mu'=(\ell+1,1,\cdots, 1), \, (\mu"=(\ell,1,\cdots, 1))$ with
 height  $k$ ($k+1)$.
Then the equality is verified by expanding  the
determinant associated to $T_{\mu"}(\lambda)$ with respect to
the first row.  

%
%----------------------
%
It remains to show the consistency of eq. (\ref{skew}) with 
eqs. (\ref{rec1c1}) ($k=n+1$ and $j=1$) and (\ref{mostfusion}).

First, consider $k=n+1$ and $j=1$ of eq.(\ref{rec1c1}). 
By the recursion argument in the above,
the formal extension $S^{(\ell)}_{n+2, 1}(\lambda)$,
defined by the rhs of eq.(\ref{skew}),  must be zero.
This is directly seen from the second expression (\ref{qJTasym}), as
$T^{(a)}_1(\lambda) =0, \, (a\ge n+2)$.

Next we deal with eq.(\ref{mostfusion}).
To be precise we will show

\begin{equation}
S^{(m+n+1)}_{i,j} (\lambda) =T^{(1)}_{m+n+1} (\lambda q^{(m+n)/2}) \delta_{i,j}
\quad (=(-1)^n \delta_{i,j}.
\label{cohomo2}
\end{equation}

With slight re-definition of index $j$ and the application of 
the relation (\ref{red2c2d}),
this is converted into an equivalent form,
\begin{equation}
S^{(m+j+1)}_{i,1} (\lambda) =T^{(1)}_{m+n+1}(\lambda q^{(m+n)/2}) 
 \delta_{i+j, n+1} \qquad (0 \le j \le n).
 \label{cohomo}
\end{equation}
We shall divide argument into two cases, $j \ne 0$ and $j=0$. \pn
For  $j = 0$,  we apply eq.(\ref{skew}) to $ S^{(m+1)}_{i,1} (\lambda)$.
Then we find from the formula (\ref{qJTsym}) that the all elements
of the first row are zero unless $i=n+1$. 
This comes from eq.(\ref{szero}), 
$T_{m+j'}(\lambda)=0$ for $1\le j' \le n$.
For $i=n+1$, We expand the determinant with respect to the last column.
From the  normalization in eq.(\ref{skew}) and $T^{(1)}_0(\lambda)=1$ , 
it follows $ S^{(m+1)}_{n+1,1} (\lambda)=
T^{(1)}_{m+n+1}(\lambda q^{(m+n)/2}) $,
which is $(-1)^n$ by  definition. \pn
A remark is in order. 
The second form (\ref{qJTasym} ) leads  to
$S^{(m+1)}_{n+1,1} (\lambda)=
(-1)^n T^{(1)}_{m}(\lambda q^{(m-1)/2}) T^{(n+1)}_1(\lambda q^{3m+n/2})  ) $.
Then 
\begin{equation}
T^{(1)}_{m}(\lambda q^{-(n+1)/2})T^{(n+1)}_{1}(\lambda q^{m/2})
=(-1)^n T^{(1)}_{m+n+1}(\lambda)=1
\label{consistency}
\end{equation}
must hold.
Neither $T^{(1)}_{m}(\lambda)$ or  $T^{(n+1)}_{1}(\lambda)$ has poles.
Thus eq.(\ref{consistency}) asserts that they are also nonzero everywhere.
Consequently they are constant. 
We drop their $\lambda$ dependencies for the time being.
By taking the determinants of both sides of (matrix form of) 
eq.(\ref{cohomo2}), one obtains
$(T^{(n+1)}_1)^{m+n+1}=(T^{(1)}_{m+n+1})^{n+1} $. 
Being combined with eq.(\ref{consistency}), this yields,
$
(T^{(n+1)}_1)^{m}= (T^{(1)}_m)^{n+1}.
$
We conclude both  $T^{(n+1)}_1$ and $T^{(1)}_m$ are root of unity.
The former can be derived directly once if we assume (\ref{open}) and use (\ref{taun}).
One only has to evaluate $W_k$ using asymptotic forms (\ref{asy})
and (\ref {symm})
in ${\cal S}_{k+(n+1)/2} \cup {\cal S}_{k+(n-1)/2} $.
The simple manipulation  yields $\tau^{(n+1)}_k=(-1)^n$, thus 
$T^{(n+1)}_1 =1$.\pn
For $j \ne 0$, the recursion (\ref{rec1c1}) yields,
\begin{eqnarray}
S^{(m+j+1)}_{i,1}(\lambda) & =& \tau^{(i)}(\lambda q^{m+j}) S^{(m+j)}_{1,1}(\lambda)
+ S^{(m+j)}_{i+1,1}(\lambda)  \np
&=&(-1)^i T^{(i)}(\lambda q^{m+j+(i-1)/2}) T^{(1)}_{m+j}(\lambda q^{(m+j-1)/2})
+ S^{(m+j)}_{i+1,1}(\lambda)  \label{cohomo1}
\end{eqnarray}
In the present case, the first term in eq.(\ref{cohomo1}) is vanishing due to
eq.(\ref{szero}). 
Thus 
\begin{equation}
S^{(m+j+1)}_{i,1}(\lambda)=S^{(m+j)}_{i+1,1}(\lambda).
\label{simpler}
\end{equation}
We then try to prove eq.(\ref{cohomo}) by induction on the upper index,  
starting from the result for $S^{(m+1)}_{i,1}(\lambda) $ as the "initial condition".
In each induction step, however,  $i=n+1$ component is indeterminate from
eq.(\ref{simpler}).
We then use  eq.(\ref{qJTasym}) and find
$S^{(\ell)}_{n+1,1}(\lambda) = (-1)^{\ell+1} 
T^{(1)}_{\ell-1}(\lambda q^{\omega(\ell)})
T^{(n+1)}_{1}(\lambda q^{\omega '(\ell)})     $
where $\omega(\ell)$ and $\omega '(\ell)$ are some shifts 
irrelevant in the present argument.
This supplies the missing pieces.
As $\ell \ge m+2$ for the case under consideration, these are null.
We then immediately check that eq.(\ref{cohomo}) holds.

Thereby,  we prove that nontrivial Stokes multipliers are explicitly given
via formula (\ref {skew}).
The fundamental quantities $\tau^{(a)}$ s are not all independent.
Constraints (\ref{szero}) impose  complex algebraic equations among them.

We note  useful relations among $T$s, the $T-$ system.
Let $T^{(a)}_{\ell}(\lambda)=T_{\mu}(\lambda)$ for $\mu$ being a 
rectangle of height $a$ and width $\ell$.
By applying the Pl{\"u}cker relation to eq.(\ref{qJTsym}) or 
eq.(\ref{qJTasym}), one finds,
\begin{equation}
T_{\ell}^{(a)}(\lambda q^{1/2}) T_{\ell}^{(a)}(\lambda q^{-1/2})
= T_{\ell+1}^{(a)}(\lambda) T_{\ell-1}^{(a)}(\lambda) +
T_{\ell}^{(a+1)}(\lambda) T_{\ell}^{(a-1)}(\lambda), \qquad a=1, \cdots, n.
\label{tsys}
\end{equation}
The boundary conditions are
$T_{\ell<0}^{(1)}(\lambda) =T_{m+1\le \ell\le m+n}^{(1)}(\lambda) =
T_{1}^{(a<0)}(\lambda) =T_{1}^{(a>n+1)}(\lambda) =0$ and
$T^{(a)}_{\ell}(\lambda) =1$ if $a=0$ or $\ell=0$.
The last relation comes from the null dimensionality  of 
the matrix in eqs.(\ref{qJTsym}) and (\ref{qJTasym}).
The initial condition $T_{\ell<0}^{(1)}(\lambda) =0$ leads 
$T_{\ell<0}^{(a)}(\lambda) =0, \, (1 \le a \le n)$.
Similarly we have $T_{m+j}^{(a)}(\lambda) =0 \, (1 \le a, j \le n) $.
Thus the $T-$ system constitutes {\it finite} number of relations
with {\it finite} number of $T^{(a)}_{\ell}(\lambda)$.
 
Remark that the case with $n=1$ ,  $m=3$ and 
with multi-parameters has been found  in 
\cite{Sbook} by direct calculation of recursion relations.
The one parameter result has been later extended to arbitrary $m$ by 
similar argument employed here \cite{DT2}.
 
Summarizing this section, we have found  determinant representations
of Stokes multipliers.
We conveniently introduce a subset of huge hierarchy,
$T^{(a)}_{\ell}(\lambda)$
including (the most fundamental) Stokes 
multipliers $\tau^{(a)}(\lambda)$.
Then the functional relations exist among them which, in turns,
impose some restrictions on  $\tau^{(a)}(\lambda)$.

%The validity of this symmetry is checked 
% by the direct matrix calculation of the recursion relations 
%for $n=1, m \in Z$ and  $m=2$  for $n= 2$ case.

We will discuss above results in view of solvable models in the
next section. Especially we will give some follow-up
of the last sentence in the previous paragraph.

\section{Functional relations in solvable models and
Thermodynamic Bethe Ansatz}

Hereafter we assume
 $n$ odd and $m=M(n+1)$.
It is then convenient to rotate the $\lambda$  
for the differential equation (\ref{difeq}) by $\frac{\pi}{2}$.
We will use the same notation for the resultant $\lambda$.

In the first part of this section, we remind of some concrete results
in solvable models.
The second part is devoted to a rather speculative discussion of
the possibility of the application of Thermodynamic Bethe ansatz 
to the evaluation 
of Stokes multipliers.

The commuting transfer matrices play a fundamental role in  studies of
solvable lattice models and field theories \cite{Baxbook, BLZ1,BLZ2}.
The members in a commuting family share  the same physical space (
quantum space) and are parameterized by the 
(multiplicative) spectral parameter $\lambda$.
They are labeled by the auxiliary space of which trace must
be taken.
These auxiliary spaces are identified with irreducible modules
of Yangian or quantum affine Lie algebra \cite{KirReshet,Kuniba}.
For  $U_q(A^{(1)}_n)$,
there exists a irreducible module $W^{(a)}_m (\lambda), (a \le n)$ 
which is isomorphic to $m V_{\Lambda_a}$ as a classical module.
Naturally, we associate a Young diagram of $a\times m$ to this module,
and write the corresponding transfer matrix ${\cal T}^{(a)}_m(\lambda)$.
Since they are commutative, we consider 
transfer matrices on common eigenstates.
Thus we sometimes do not distinguish operators from their eigenvalues.

In the language of solvable lattice models, ${\cal T}^{(a)}_m(\lambda)$ is a 
transfer matrix of  a model obtained by the fusion procedure.
Starting from a "fundamental" model 
acting on $V_{\Lambda_1} \times V_{\Lambda_1}$,
we can recursively derive fusion models.
Their auxiliary  spaces
are constructed  from $V_{\Lambda_1}\times \cdots \times V_{\Lambda_1}$ by
applying appropriate projectors.
We utilize singular points of $R-$ matrix for
construction of these projectors.

The fusion procedure is not restricted to 
the rectangle type.
Generally, there exists a  module  $W_{\mu/\eta}(\lambda)$, or a fusion model, 
parameterized by  a skew Young diagram ${\mu/\eta}$.
We write the corresponding transfer matrix  as ${\cal T}_{\mu/\eta}(\lambda)$.

The short exact sequence of irreducible modules examined in 
\cite{Cherednik,BR, KNS1}
leads to the relation which we have seen 
in eq.(\ref{qJTsym}),

\begin{eqnarray}
{\cal T}_{\mu/\eta}(\lambda) &=&
  {\rm det }_{1\le j,k\le \mu_1} 
         ( {\cal T}^{(\mu'_j-\eta'_k-j+k)}_{1} 
            (\lambda q^{-(\mu_1' -\mu_1-\mu'_j-\eta'_k+j+k-1)/2}) )     
			 \label{TqJasym} \\
  &=&{\rm det }_{1\le j,k\le \mu_1'} 
         ( {\cal T}^{(1)}_{\mu_j-\eta_k-j+k} 
            (\lambda q^{-(\mu_1' -\mu_1+\mu_j+\eta_k-j-k+1)/2}) )
\label{TqJsym}
\end{eqnarray}
where ${\cal T}^{(1)}_{m<0}(\lambda) ={\cal T}^{(a<0)}_1(\lambda) =0$.\pn
This  assures the $T-$ system (\ref{tsys}) among ${\cal T}$ \cite{KNS1}.
The  quantum Jacobi Trudi formula plays a role in several problems in
solvable models \cite{KOS, SuzE8, SuzE7}

The conditions (\ref{szero}) also hold.
When $q=\exp(i \frac{2\pi}{m+n+1})$, the truncation of
the space happens due to  quantum group symmetry.
In view of solvable lattice models, it corresponds
to the situation that some projectors are vanishing
and fusion paths in local variables  are lost. 
Consequently, no local variables can be adjacent
and transfer matrices vanish\cite{BR},
\begin{equation}
{\cal T}^{(a)}_{m+1}(\lambda) ={\cal T}^{(a)}_{m+2}(\lambda) 
=\cdots={\cal T}^{(a)}_{m+n}(\lambda) =0.
\label{truncation}
\end{equation}

The normalizations of ${\cal T}^{(a)}_0(\lambda), 
{\cal T}^{(0)}_{\ell}(\lambda) $ and ${\cal T}^{(n+1)}_{\ell}(\lambda) $ depend on
choice of quantum space.
We have not yet found the description of the  
quantum space yielding ${\cal T}^{(a)}_0(\lambda)=
{\cal T}^{(0)}_{\ell}={\cal T}^{(n+1)}_{\ell}=1$, 
except for $n=1$.
We assume the existence of it for general $n$.

Then, on the corresponding space, the fusion transfer matrices share
same  functional relations with Stokes multipliers.

Below we use the same  symbol $T^{(a)}_{\ell} (\lambda)$ for these two cases.

Thanks to the (quantum group) reduction,the $T-$ system closes
within a finite set of unknowns.

The  solution to  functional relations, however, is not unique.
Additional information on the analyticity of 
$T^{(a)}_{\ell}(\lambda)$ in the complex $\lambda$  are required
for the uniqueness.\pn
We start from the lattice model.
Then we take the scaling limit (or the field theoretical limit) which has been
 discussed in several literatures.
In the present context, we refer to \cite{PAPBN}.
Note that our transfer matrices are not mere
sums of products of Boltzmann weights 
but are imposed overall renormalizations due to
${\cal T}^{(a)}_0(\lambda)=
{\cal T}^{(0)}_{\ell}={\cal T}^{(n+1)}_{\ell}=1$.

Let us define an "additive" spectral parameter $u$ by  
$\lambda =\exp(\frac{\pi u}{(m+n+1)})$.
We also introduce 
$$
Y^{(a)}_{\ell}(u) :=
 \frac{T^{(a)}_{\ell+1}(\lambda(u))T^{(a)}_{\ell-1}(\lambda(u)) }
 {T^{(a+1)}_{\ell}(\lambda(u))T^{(a-1)}_{\ell}(\lambda(u)) }
 \quad (1\le a \le n, 1\le \ell \le m).
$$
Then  eq.(\ref{tsys}) reads 
\begin{equation}
Y^{(a)}_{\ell}(u+i) Y^{(a)}_{\ell}(u-i)=
\frac{(1+Y^{(a)}_{\ell+1}(u) )(1+Y^{(a)}_{\ell-1}(u) )}
{ (1+(Y^{(a+1)}_{\ell}(u))^{-1} )(1+(Y^{(a-1)}_{\ell}(u))^{-1} )},
\label{ysys}
\end{equation}
where  $(Y^{(0)}_{\ell}(u))^{-1}=Y^{(a)}_{m}(u)=0$ by  definition.

Remember that that Boltzmann weights are regular functions of $u$.
Then, apart from some renormalization factors mentioned above,
$T^{(a)}_{\ell}(\lambda( u))$ 
must not be singular in the strip $\Im u \in [-1, 1]$ but
may possess zeros in general.

Suppose that $T^{(a)}_{\ell}(\lambda( u))$ has 
finitely many distinct zeros in the strip  
$\{u^{(a)}_{\ell, k} \}$  which depend on  eigenstates.
They are assumed to be off the lines  $\Im u =\pm 1$.
Then  eq. (\ref{ysys}) can be transformed into coupled nonlinear 
integral equations
by the  standard trick\cite{KP, Klu92, BLZ1, BLZ2, JKSfusion, KSS98}.

The both sides of eq. (\ref{ysys}) are analytic with finitely many distinct zeros
and have constant asymptotics.
Then, applying  Cauchy's theorem, we have
\begin{eqnarray}
& &\log Y^{(a)}_{\ell} (u) =
\log {\cal Z}^{(a)}_{\ell}(u) \np
  & & +\sum_{b=1}^a  \sum_{r=1}^{m} 
   \int_{-\infty}^{\infty} K(u-u')
    \log   \frac{(1+Y^{(a)}_{\ell+1}(u') )(1+Y^{(a)}_{\ell-1}(u') )}
{ (1+(Y^{(a+1)}_{\ell}(u'))^{-1} )(1+(Y^{(a-1)}_{\ell}(u'))^{-1} )}  du' + C^{(a)}_\ell
\label{tba1}
\end{eqnarray}
where $C^{(a)}_\ell$ is a "integral" constant fixed by 
comparing asymptotic values of both sides.

$ {\cal Z}^{(a)}_{\ell}(u) $ signifies,
\begin{eqnarray}
{\cal Z}^{(a)}_{\ell}(u)&=&
\frac{Z^{(a)}_{\ell+1}(u)  Z^{(a)}_{\ell-1}(u) }
   { Z^{(a+1)}_{\ell}(u) Z^{(a-1)}_{\ell}(u) }     \np
Z^{(a)}_{\ell}(u)&=&\prod_j 
 ( \tanh \frac{\pi}{4} (u-w^{(a)}_{\ell, j}) )^{\epsilon_{\ell,j}   }  \times
   \prod_k \tanh \frac{\pi}{4} (u-u^{(a)}_{\ell,k}) .
   \label{zdef}
\end{eqnarray}
and  $Z^{(0)}_{\ell}(u)=Z^{(n+1)}_{\ell}(u)=Z^{(a)}_{0}(u)=1$.

Here $\{w^{(a)}_{\ell, j}) \}$ is the joint set of 
zeros ($ \epsilon_{\ell,j}=1 $)
and singularities ($ \epsilon_{\ell,j}=-1 $) of 
$T^{(a)}_{\ell}(u)$ in $\Im u \in [-1,1]$ due to
the above normalization. 
They stem from common factors in  Boltzmann weights, 
thus they are zeros or singularities  of  order $N$ ($N=$ system size).
Here we label them disregarding of their multiplicities.

The kernel function is easily written in the Fourier transformed form.
We define the Fourier transformation $\widehat{f}[k]$ of a function $f(u)$ by
$$
f(u) = \frac{1}{2 \pi} \int_{-\infty}^{\infty} \widehat{f}[k] e^{iku} du,
\quad 
\widehat{f}[k] = \int_{-\infty}^{\infty} f(u) e^{iku} du,
$$
then 
$$
\widehat{K}[k]= \frac{1}{2 \cosh k}.
$$

Eq. (\ref{tba1}) might be represented by the following form,
in analogy with thermodynamic Bethe ansatz equation in CFT,

\begin{eqnarray}
\log \widetilde{Y^{(a)}_{\ell}} (u) &=& 
  \sum_{r=1}^{m} \int_{-\infty}^{\infty}
         {A}^{\ell,r}(u-u') \log {\cal Z}^{(a)}_r (u') du'\np
  & & +\sum_{b=1}^a  \sum_{r=1}^{m} 
   \int_{-\infty}^{\infty}  K_{a,b}^{\ell,r} (u-u') \log(1+(Y^{(b)}_{r} (u'))^{-1})  du' +D^{(a)}_\ell
\label{tba2}
\end{eqnarray}
where  $D^{(a)}_\ell$ is also some  "integral" constant
and $\widetilde{Y^{(a)}_{\ell}} (u) =Y^{(a)}_{\ell} (u)/Z^{(a)}_{\ell} (u)$.

The Fourier transformations of the kernel functions read,
\begin{eqnarray*}
\widehat{K}_{a,b}^{\ell,r}[k]&=& \widehat{A}^{\ell,r}[k] \widehat{M}_{a,b}[k] \\
\widehat{A}^{\ell,r}[k]&=& 
\frac{\sinh({\rm min}({\ell,r} )k)  \sinh((m-{\rm max}({\ell,r}))k)     }
{\sinh(mk) \sinh(k)}   \\
\widehat{M}_{a,b}[k]&=& 2 \cosh k (\delta_{a,b} -\frac{I_{a,b}}{ 2 \cosh k })\\
\end{eqnarray*}
and $ I_{a,b}=1$ if they are on  adjacent nodes of Dynkin
diagram for $A_n$ and  $ I_{a,b}=0$ otherwise.

At this stage, one performs the field theoretical limit ,
system size $\rightarrow \infty$, lattice spacing  $\rightarrow 0$,
and sending elliptic nome $\rightarrow 0$ under some fine-tuning condition.
Precisely speaking, this condition depends on the regime 
of the lattice model. 
We shall skip that detail here.
We refer \cite{PAPBN} for detail in the case of $A^{(1)}_1$.

Then it can be shown that  $\{w^{(a)}_{\ell, j} \}$
accumulate in $\infty$ and reproduce "momentum" term in eq.(\ref{tba2}) from
the first term \cite{DT1,BLZ1,BLZ2, BLZ3, PAPBN},

\begin{eqnarray}
\log \widetilde{Y^{(a)}_{\ell}} (u) &=& M^{(a)}_{\ell} \exp( \frac{\pi}{m}  u) +
  \sum_{r=1}^{m}  \int_{-\infty}^{\infty}
       {A}^{\ell,r}(u-u') \log {\cal Z'}^{(a)}_r (u') du'   \np
  & & +\sum_{b=1}^a  \sum_{r=1}^{m} 
   \int_{-\infty}^{\infty} 
      K_{a,b}^{\ell,r} (u-u') \log(1+(Y^{(b)}_{r} (u'))^{-1})  du' 
   +D^{(a)}_\ell.
\label{tba3}
\end{eqnarray}
${\cal Z'}^{(a)}_r (u')$ consists only of $\{u^{(a)}_{\ell, k} \}$
in (\ref{zdef}).

The rapidity $\theta$ in \cite{DT1} is
 denoted by $\frac{\pi}{m}  u$ here.\pn
The resultant coupled integral equations 
can be recursively solved and yield a unique set of solutions to
$Y^{(a)}_{\ell}(u)$, and then to $T^{(a)}_{\ell}(\lambda(u))$
for given $\{u^{(a)}_{\ell, k} \}$ and  $ M^{(a)}_{\ell}$.
Thus one must adopt appropriate choice of these parameters  to reproduce
proper $T^{(a)}_{\ell}(\lambda(u))$ including $\tau^{(a)}(\lambda)$.
Conversely, if two sets of functions satisfy the same relation (\ref{tsys}) and 
share the same $\{u^{(a)}_{\ell, k} \}$ and $ M^{(a)}_{\ell}$, 
then they are identical in the strip.

For solvable lattice models, the case studies on $U_q(A^{(1)}_{1,2,3})$ 
indicate that  $T^{(a)}_{\ell} (\lambda(u))$ does not possess zeros in the strip 
$\Im u \in [-1, 1]$ when acting on the largest eigenvalue sector.
That is, $\{u^{(a)}_{\ell, k} \}=\phi, a=1,\cdots n$ 
for all $p\le m$. \pn
This seems to be also the case with Stokes multipliers.
Actually this is the case for $n=1$.
That is, the proper Stokes multipliers are reproduced by  
$\{u^{(a)}_{\ell, k} \}=\phi, a=1,\cdots n$ 
for all $p\le m$ but different choice of $M^{(a)}_{\ell}$
 from lattice models
or field theoretic model.
Let me just present some argument for $n \ge 2$ case.
Assume that the solution to eq. (\ref{consistency})
 is simple, $T^{(n+1)}_1=T^{(1)}_m =1$.
As remarked above, the former equality
is a direct consequence of (\ref{open}) and (\ref{taun}).
From our "boundary conditions", the $T-$ system is then
invariant under the simultaneous transformations
$(a, \ell) \rightarrow (n+1-a, m-\ell) $.
This implies $T^{(a)}_{\ell} (\lambda) =T^{(n+1-a)}_{m-\ell} (\lambda)$.
By $\pm \theta^{(a)}_{\ell}$, we mean  the imaginary part of zero 
of $T^{(a)}_{\ell} (\lambda)$.
Then a sum rule holds, 
$$
\theta^{(a)}_{\ell} + \theta^{(n+1-a)}_{m- \ell}=m+n+1.
$$
The simplest solution, $\theta^{(a)}_{\ell}=a+\ell$,
 is actually correct for  $n=a=1$.
Postulating this solution, we conclude  
$\{u^{(a)}_{\ell, k} \}=\phi, a=1,\cdots n$ 
for all $p\le m$ also for Stokes multipliers.

We do not expect ill-behavior for 
small angle sector of $\lambda$;
 $\tau^{(a)}(\lambda)$ might not be vanishing  in the sector, or otherwise 
it harms linear dependency of FSS.
The above  choice of $\theta^{(a)}_{ \ell}$ is 
consistent with this expectation.

Admitting these assumptions, we have a conjecture
\begin{cj}
The Stokes multipliers of eq.(\ref{difeq}) are given by
"vacuum expectation values" of fusion transfer matrices related
to $U_q(A^{(1)}_n)$.
Thus they are evaluated from eq.(\ref{tba3}) by forgetting the second 
 term  in the rhs .
\end{cj}
The parameters $M^{(a)}_{\ell}, D^{(a)}_{\ell}$ must be tuned correctly as in
\cite{DT1} for $n=1$.

To be precise, overall normalization of $\lambda$ for the identification
is not fixed by  functional relations only.
For $n=1$ the determination of this factor is quite straightforward  \cite{DT1,DT2},
 due to the fact that  only one subdominant
solution exists in each ${\cal S}_j$.
The spectral determinant is then identified with the 
fusion Stokes multiplier which connects 
the subdominant solution on the negative real axis to
the dominant solution on the positive real axis.
Then standard WKB arguments yield the identification of
parameters, especially $M^{(1)}_{\ell}$ and the
 normalization of $\lambda$.

This can not be generalized straightforwardly
for $n \ge 2$. 
Let me just comment on two fundamental problems.
First, the meaning of eigenvalue problem is not necessary 
clear for higher order differential equations.
The characterization of the eigenspace (it is the Hilbert space for
$n=1$) is not obvious.
Second, more technically,
there are several subdominant solutions in each sector.
We hope to clarify these issues in future publications.

Finally we comment on the (1,1) component of eq.(\ref{stokes0}).
\begin{equation}
y_k = \tau^{(1)}_k y_{k+1} + \tau^{(2)}_k y_{k+2}+ 
\cdots + \tau^{(n+1)}_k y_{k+n+1}.
\label{spectral}
\end{equation}
The equation coincides with the Baxter's $T-Q$ relation for
$n=1$ by putting $x=0$ and writing $y_k(x=0) =Q(\lambda q^{k+{\rm const.}})$.
For general $n$, this coincides with the "spectral curve" equation
argued in \cite{FF} or the characteristic equation in 
the quantum separation of variables \cite{Sklyanin}.

\section{discussion}
We have seen that Stokes multipliers
associated to a $n+1$ th differential equation
satisfy the same recursion relations with
fusion transfer matrices related to $U_q(A^{(1)}_n)$.
Under the assumption on analyticity of some functions in the strip, 
we conjecture  thermodynamic Bethe ansatz type equations eq.(\ref{tba3})
which determine Stokes multipliers.
The assumption , however,  needs extensive numerical check, which 
we hope to report in the near future. 

The deformation parameter  
$q=\exp(i\frac{2\pi}{m+n+1})$, which arises naturally in the present context,
 has a concrete meaning in the solvable models.  
The parameter $m$ specifies the level of dominant integral
weight of the local variables in the lattice model\cite{JMO}. 
Then the denominator of the exponent of $q$ is of the form,
 level + dual Coxeter number.
This combination also appears as  the deformation parameter of a
solvable model based on other affine Lie algebras \cite{rDJO}.
We thus expect a possible extension of the present study to
other type of potential terms related to other affine Lie algebras.

\section*{Acknowledgments}
The author thanks Y. Takei  for discussions and comments.
He also thanks A. Voros for comments and calling his attention to
references \cite{V6,V7, Olver}.

\section*{Appendix}
We present the argument leading to eq.(\ref{open}).

In terms of the vector $u:=(y, \partial y, \cdots, \partial^n y)$,
eq.(\ref{difp}) reads,
$$
\frac{d}{dx} u = A(x) u =
\begin{pmatrix}
       0,&   1,& 0,&  \cdots,&    0 \\
       0,&   0,& 1,&  \cdots,&    0 \\
  \vdots &     &   &         & \vdots\\
       0,&   0,& 0,&  \cdots,&    1 \\
   P(x), &   0,& 0,&  \cdots,&    0 
\end{pmatrix}
$$
We define a new variable $\xi$ by $\xi^{n+1}=x$.
Put $u=M T w$ such that  $M={\rm diag}(1, \xi^m, \xi^{2m}, \cdots )$
and that $T$ diagonalizes a matrix of the form $A(x)$ in the
above with  $P(x)=1$.
Then the equivalent matrix equation is 
written in the form
\begin{equation}
\frac{d}{d\xi} w = {\xi}^{m+n} B(\xi) w =\xi^{m+n} \sum_{j=0} B_j {\xi}^{-j} w ,
\label{lin2}
\end{equation}
and $B_0 ={\rm diag}(\exp(i \frac{2\pi n}{n+1}), 
\exp(i \frac{2\pi (n-1)}{n+1}), 
 \cdots, 1 )$.

Following \cite{HS, Sbook}, we assume $w$ in the form,
$$
w=\begin{pmatrix}
      p_1(\xi) \\
	  p_2(\xi) \\
	  \vdots   \\
      p_{n+1}(\xi)
\end{pmatrix}
  \exp( \int^{\xi} \eta^{m+n} \gamma(\eta) d\eta ),
$$
and $p_1(\xi)=1$.  \pn
Then the substitution of the above form into eq. (\ref{lin2}) yields,
\begin{equation}
\xi^{-m-n} \frac{d}{d\xi} p_j = \beta_{j,1}+
(\beta_{j,j}-\beta_{1,1}) p_j   -
\sum_{i \ne 1} \beta_{1,i} p_i p_j +
\sum_{k \ne 1, j} \beta_{j,k} p_k , 
\label{lin3}
\end{equation}
where $\beta_{j,k}$ denotes the $(i,j)$ th component of $B(\xi)$ 
and $j \ge 2$. Note that $\beta_{j,k}={\cal O}(1/\xi)$ if
$j \ne k$.

Immediately seen, $\widehat{p_{j}}(\xi) =\sum_{N=1} p_{j,N} \xi^{-N}$ 
is a formal solution; one can recursively determine $p_{j,N}$
by (\ref{lin3}).

We define $h_0 = \beta_{n+1,n+1}(\infty) - \beta_{1,1}(\infty)$, and 
the sector ${\cal S}$ in the $\xi-$ plane  by
$$
|{\rm arg} h_0 + (m+n+1) {\rm arg} \xi | \le \frac{3}{2}\pi ,
\quad |\xi| \ge \Omega
$$
for a fixed positive $\Omega$ .
By ${\cal D}_r$ we mean the domain in $(a_1, \cdots, a_m)$ space
such that
$$
|a_1|^2+ \cdots |a_m|^2 < r
$$
for some  positive $r$.
Then we naturally generalize the result in \cite{HS, Sbook}.
\begin{cj}
one can construct  $\widehat{p_{j,r}}(\xi) $ so that $\widehat{p_{j,r}}(\xi) $
is holomorphic with respect to $(\xi, a_1,  \cdots, a_m)$ in
${\cal S} \times {\cal D}_r$ and $\partial^k \widehat{ p_{j,r}}(\xi), (k=0,1,\cdots)$
 admits the uniformly asymptotic expansions,
$$
\partial^k \widehat{p_{j,r}}(\xi) \sim  \partial^k  \widehat{p_{j}}(\xi)
$$
for $(a_1, \cdots, a_m) \in {\cal D}_r$ as $ \xi$ tends to infinity
in ${\cal S}$.
\end{cj}

Let us present some argument supporting this (by no means a proof).
We put $p_j = q_j+ \widehat{p_{j,r}}(\xi) $ and 
inserting this into (\ref{lin3}), 
\begin{eqnarray*}
\xi^{-m-n} \frac{d}{d\xi} q_j &=& C_{j,1}+
         C_{j,k} q_j  -  \sum_{i \ne 1} \beta_{1,i} q_i q_j 
\quad (j \ge 2),  \\
C_{j,1}&=&\beta_{j,1}+(\beta_{j,j}-\beta_{1,1}) \widehat{p_{j,r}}(\xi) 
              +\sum_{k \ne 1,j} \beta_{j,k} \widehat{p_{k,r}}(\xi) \\
       & & - \sum_{i\ge 2} \beta_{1,i} \widehat{p_{i,r}}(\xi) \widehat{p_{j,r}}(\xi)
	    -\xi^{-m-n}\frac{d}{d\xi} \widehat{p_{j,r}}(\xi) \quad(j \ge 2),  \\
C_{j,k}&=&\beta_{j,k}(1-\delta_{j,k})
+ \delta_{j,k} \{ (\beta_{j,j}-\beta_{1,1}  )  -\sum_{i \ge 2} 
  \beta_{1,i} \widehat{p_{i,r}}(\xi)  \}   \\
 &-& \beta_{1,k} \widehat{p_{j,r}}(\xi) , \quad(j,k \ge 2). \\
\end{eqnarray*}

The formal solution is 
\begin{equation}
q_j(\xi) = \int^{\xi} \eta^{m+n} 
 \Bigl (
  C_{j,1} + \phi_{j,k} q_k + \sum_{i \ne 1} \beta_{1,i} q_i q_j 
    \Bigr )  \times
 \exp(\frac{h_0^{(j)} }{m+n+1}( \xi^{m+n+1}- \eta^{m+n+1}) )  d \eta.
\label{formal}
\end{equation} 
where we set $C_{j,k} = h_0^{(j)}\delta_{j,k}+\phi_{j,k}$
and $h_0^{(j)}=(n+1) (\exp(2 \pi i\frac{n+1-j}{n+1})-
\exp(2 \pi i\frac{n}{n+1}) )$.
Now the coefficients in the bracket of the integrand are less than order
$\frac{1}{\xi}$.

We are interested in eq.(\ref{asy}). 
For this, it may be enough to treat $j=n+1$.
Assume there exists  appropriate choice of $M$, then  Lemma 3 of \cite{HS}
also applies in this case. 
The  integral in the rhs of (\ref{formal}) can be thus bounded from above ,
$|q_{n+1}| \le {\cal O}(|\xi|^{-1})$ in a certain domain provided that
\begin{equation}
|{\rm arg} h_0^{(n+1)} + (m+n+1){\rm arg} \xi | < \frac{3 \pi}{2} -\epsilon
\label{range}
\end{equation}
where $\epsilon$ is  positive and small.
By definition, $ h_0^{(n+1)}=h_0$, thus in ${\cal S}$,
$q_{n+1}$ tends to zero as $|\xi| \rightarrow \infty$.
Hence the solution converges to the asymptotic expansion form.
It may be shown that the limit does not depend on $r$ as in
\cite{HS, Sbook}.
The condition (\ref{range}) coincides, when written in terms of $x$, 
 with (\ref{open}).

%
%---------------------------------------------------
%

\end{document}